\documentclass[a4paper,11pt]{article}

\usepackage{adjustbox}

\usepackage{amsmath}
\usepackage{amssymb}

\usepackage{graphicx}
\usepackage{color}
\usepackage{float}

\title{Could random dynamics derive quantum mechanics via the weak value?}

\author{Holger Bech Nielsen\footnote{Speaker at the workshop 
``What comes beyond the Standard Models'' in Bled 2024.}\\
Niels Bohr Institute, University of Copenhagen \\
Blegdamsvej 17, Copenhagen $\O$, Denmark \\
and \\
  Keiichi Nagao\\
  Faculty of Education, Ibaraki University \\
Bunkyo 2-1-1, Mito 310-8512 Japan }

\date{``Bled''   , July , 2024}

\vspace{-1mm}

\begin{document}

\maketitle

\begin{abstract}
We argue that we could make  a scenario of deriving quantum mechanics, 
as a random dynamics project, in the sense of it being almost unavoidable. 
The basic idea is based on the weak value formulation. 

\end{abstract}

\noindent Keywords: non-Hermitian Hamiltonian, inflation, weak value

\noindent PACS: 11.10.Ef, 01.55 +b, 98.80 Qc.

\section{Introduction}\label{s:intro}

Random dynamics - see e.g \cite{RDsweden, RDTeitelboim,RDOverbye, RDAstri, RDColin} - is a project of calculations in which one or several laws of nature 
are {\it not assumed}, but hoped to be {\it derived}. 
Indeed it would be very nice if we could realize a theory as fundamental as possible. 
More fundamental theories should have less conditions supposed at first. 
For example, we are usually accustomed to using real actions in many kinds of theories, 
but using real actions by itself means imposing on actions one common restriction 
that each action has to be real. 
If we hope for more fundamental theories, we have to be free of such a restriction. 
Based on such insight, the complex action theory (CAT) was initiated~\cite{Bled2006}. 
In the CAT, not only many falsifiable 
predictions~\cite{Bled2006,Nielsen:2008cm,Nielsen:2007ak,Nielsen:2005ub} 
but also various topics such as the Higgs mass\cite{Nielsen:2007mj}, quantum mechanical philosophy\cite{newer1,Vaxjo2009,newer2}, 
some fine-tuning problems\cite{Nielsen2010qq,degenerate}, black holes\cite{Nielsen2009hq}, 
de Broglie-Bohm particles and a cut-off in loop diagrams\cite{Bled2010B}, 
the complex coordinate and momentum formalism\cite{Nagao:2011za}, 
the momentum relation~\cite{Nagao:2011is,Nagao:2013eda}, 
and the harmonic oscillator model~\cite{Nagao:2019dew} were studied. 

Even if a given action is complex, which means a given Hamiltonian is 
non-normal\footnote{The Hamiltonian $H$ is 
generically non-normal, so it is 
not restricted to the class of PT-symmetric non-Hermitian Hamiltonians that were 
studied in Refs.\cite{Bender:1998ke,Bender:1998gh,Mostafazadeh_CPT_ip_2002,Mostafazadeh_CPT_ip_2003,Bender:2011ke}.}, 
we could effectively obtain a Hermitian Hamiltonian after a long time development\cite{Nagao:2010xu}. 
This is a very nice property by which the CAT could be viable. 
Here, to say that more accurately,  we note that we need to 
introduce a modified inner product\cite{Geyer, Nagao:2010xu} 
such that a given non-normal Hamiltonian becomes normal with regard to it. 
In this article, however, we ignore it for simplicity because we do not need it 
for the main purpose of this article. 
We have two types of the CAT. 
One type is the future-not-included theory, where 
only a past state $| i (t_i) \rangle$ at the initial time $t_i$ is given. 
The other type is the future-included theory, where not only a past state $| i (t_i) \rangle$ 
but also a future state $| f (t_f) \rangle$ at the final time $t_f$ is given. 
Even though the future-not-included CAT has many intriguing properties\cite{Nagao:2013eda}, 
it is not favored from a point of view of Feynman path integral~\cite{Nagao:2017ecx}. 
Therefore, we think that the future-included theory is more important than 
the future-not-included theory.

In the future-included theory\footnote{Recently, in the context of quantum gravity, an example of the future-included CAT was derived based on the group field theory coupled to a scalar field, 
and its possible implication was discussed\cite{Liu:2023pok}.}, 
we have studied the construction of the so called 
weak value~\cite{AAV,generalized_two-state_vector_formalism} 
of an operator $O$ that is the ratio 
which would look in the Heisenberg representation
\begin{eqnarray}
      O_\text{weak value}(t) &=& \frac{\langle f |O(t)|i\rangle}{\langle f | i \rangle}, \label{O_weakvalue}
\end{eqnarray}
where  $|i\rangle$ and $|f\rangle$ are an initial state at the initial time $t_i$ and a final state 
at the final time $t_f$, respectively.
The weak value has been investigated in the real action theory (RAT). 
For details, see Ref.\cite{review_wv} and references therein.
This is the expression, which we used and suggested as giving an average 
useful in our complex action theory\cite{Bled2006}. 
Indeed, regarding it as an expectation value leads to obtaining the Heisenberg equation, Ehrenfest's theorem, 
and a conserved probability current density\cite{Nagao:2012mj,Nagao:2012ye}.

Thus the weak value has nice properties, but it has a serious problem: it is generally not real but 
complex even for Hermitian $O$, 
though it has to be real if it is expected to work as an observable. 
To resolve this problem, in Refs.\cite{Nagao:2015bya,Nagao:2017cpl}, 
we proposed a theorem that states that, 
provided that an operator $O$ is Hermitian, 
the weak value of $O$ becomes real and time-develops under an effectively obtained 
Hermitian Hamiltonian for the past and future states selected 
such that the absolute value of the transition amplitude from the past state 
to the future state is maximized. We call this way of thinking the maximization principle. 
We proved this theorem in the case of non-normal 
Hamiltonians $H$\cite{Nagao:2015bya} and in the RAT~\cite{Nagao:2017cpl}. 
The maximization principle is reviewed in Refs.\cite{Nagao:2017book,Nagao:2017ztx}. 
We also found, in the periodic CAT, that a variant type of the maximization principle 
can select the period\cite{Nagao:2022rap}.

The weak value in the Heisenberg representation (\ref{O_weakvalue}) is expressed 
better in the Schr{\"o}dinger representation 
\begin{eqnarray}
O_\text{weak value}(t) &=& \frac{\langle f |\exp(- \frac{i}{\hbar} H(t_f-t))O\exp(-\frac{i}{\hbar} H(t-t_i))|i\rangle}{
    \langle f |\exp(- \frac{i}{\hbar}  H(t_f-t_i))|i\rangle}, 
\end{eqnarray}
where $H$ is a given non-normal Hamiltonian, and the states $|i (t)\rangle$ and $|f (t)\rangle$ are supposed to 
time-develop according to the Schr{\"o}dinger equations 
$i\hbar \frac{d}{dt} | i (t) \rangle = H| i (t) \rangle$ and  $i\hbar \frac{d}{dt} | f (t) \rangle = H^\dag | f (t) \rangle$, respectively. 
Rewritten in functional integral formulation, this weak value 
becomes, say with the understood boundary values at the initial and final states, 
\begin{eqnarray}
O_\text{weak value}(t) &=&\frac{\int O(t)\exp(\frac{i}{\hbar} S[\text{hist}]){\cal D}\text{hist}}
       {\int \exp (\frac{i}{\hbar} S[\text{hist}]){\cal D}\text{hist}}, 
\end{eqnarray}
where $\text{hist}$ stands for the history of the fields and ${\cal D}\text{hist}$ is 
the functional integration measure.

Now the main point of this manuscript is to call attention to that
{\it if we took the action $S[\text{hist}]$ to be purely imaginary, so
  that $iS[\text{hist}]$ was purely real, then the weak value 
in the functional integral formulation could be considered an ordinary probability formula 
for the average of the variable $O(t)$.}
If we let 
\begin{eqnarray}
  \Sigma |O(t) = O'\rangle \langle O(t)= O'|
\end{eqnarray}
be the sum over a set of products of eigenstates with the eigenvalues of $O(t)$ being
$O'$, then this operator would be a projection operator on the eigenstates of $O(t)$. 
For example, in the Heisenberg picture, 
the ``probability'' for the eigenvalue of $O(t)$ being $O'$, 
$P_{O(t)=O'}$, would be
\begin{eqnarray}
P_{O(t)=O'} &=& \Sigma  \langle f |O(t) = O'\rangle \langle O(t)= O'|i\rangle. 
\end{eqnarray}
For the weak value being a good replacement for the usual quantum mechanics
average of an operator formula, these weights should be positive or zero.
We have not yet shown that, but we made some theorems 
about reality\cite{Nagao:2015bya,Nagao:2017cpl}, which we explained above. 
To deduce that this distribution should at least be real is not obvious at all to start with. 
If we took $i$ to be real as our playing assumption which is of course not true, 
then we could ensure the reality easily for any Hermitian $O'$.  
So, in this absurd case, the weak value would look like a probability
formula, except that the probabilities could be negative. 
But the crux of the matter is that {\it the weak value formally looks like a probability
distribution.} 
So, if we achieved some speculative model providing us some probability distribution 
- from some graph theory or whatever -, we could claim that now we want to write that 
as a weak value theory formally, and then we could play in the CAT to describe our world  
under the wild assumption that $S$ is purely imaginary. 
We consider some system of dynamical variables such as fields 
that makes up a complete set of variables, and have some theory for their distribution. 
Even though our theory has no quantum mechanics, we can just declare 
the exponentiated purely imaginary action $\exp(\frac{i}{\hbar}S[\text{hist}])$ 
to give distribution in the quantum mechanics lacking theory we start with.
So, {\it if one could invent a model-speculation that could provide complex numbers 
to come into the probability, then we might be able to derive  
the weak value quantum theory. }
Of course, it looks too wild to hope to find such a scenario. 
But, if one could, it would be using the weak value to ``derive'' quantum
mechanics, and one needs strongly some derivation from very little of
quantum mechanics in random dynamics.

This manuscript is organized as follows. 
In section 2, we explain some mild assumptions 
that we make in a general model based on the random dynamics, and define a specific ``action'' $S_P[q]$.   
In section 3 we give a phenomenological example of the ``action'' $S_P[q]$ and argue that 
a favorable path could fit the cosmology.  
In section 4 we introduce a beating ``clock'' in a subsystem, and argue 
that, by the beating ``clock'', a kind of interference could be caused in the other subsystem. 
In section 5, after briefly explaining the CAT and the weak value, we 
argue that the effect of the ``clock'' could give us the weak value complex path integral. 
In addition, we discuss how we could add a phase to the logarithm of the action so that 
our general formalism matches the weak value expression. 
Section 6 is devoted to discussion.

\vspace{1cm}

\section{Formulation of general model via random dynamics}

We start from the formulation of general model via random dynamics. 
We do not put in say quantum mechanics, but do not exclude it either. 
Such a formulation is a rather empty framework as described below: 
\begin{itemize}
\item{\bf Variables and time}

A lot of dynamical variables are described for short as just one
\begin{eqnarray}
q &=& (q_1, q_2, ..., q_N ), 
\end{eqnarray}
which is taken as functions of time 
\begin{eqnarray}
q(t) &=& (q_1(t), q_2(t),...,      q_N(t)). 
\end{eqnarray}
Here $N$ may be infinite or finite or even the $q$'s could be fields.  
$N$ could be $\text{Card}({\bf R})$. 

\end{itemize}

\begin{itemize}
  \item{\bf Probability distribution for paths}
    
    We assume that details of the theory should give us a functional probability
    distribution $P$ on the space of all histories $q: \{ \text{time} \} \rightarrow
    {\bf R}^N$ thinkable:
    \begin{eqnarray}
      P[q] &=& \hbox{probability distribution on sets of functions $q$}.
      \end{eqnarray}
\end{itemize}
$P[q]$ gives probabilities for paths $q.$

\vspace{1cm}

\begin{figure}
\centering
\includegraphics[scale=0.6]{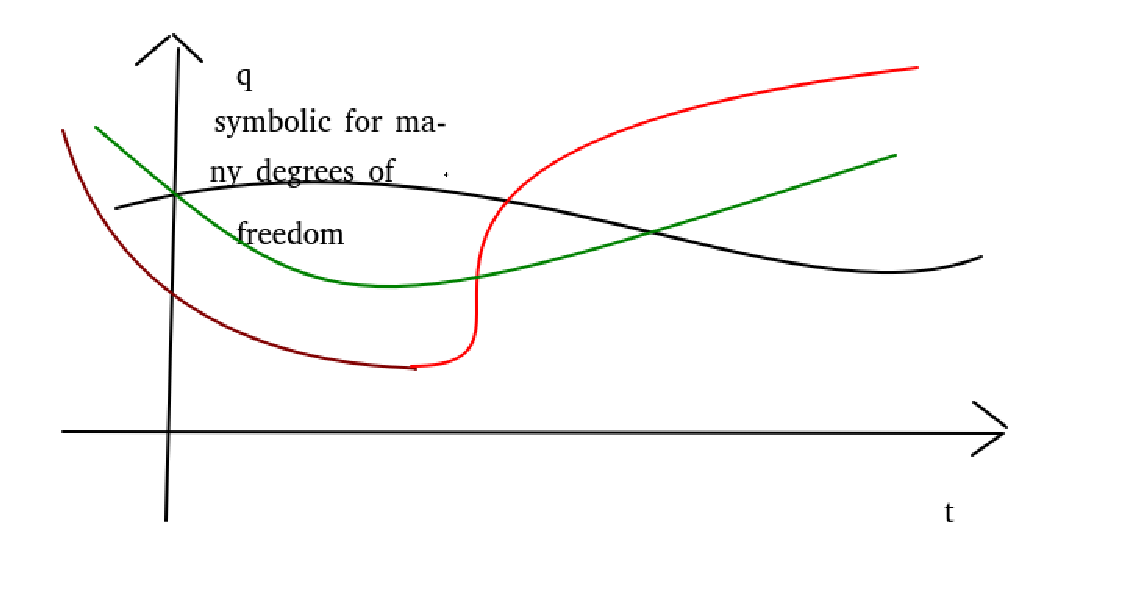}
\caption{Various paths}
\label{figure:various_paths}
\end{figure}

Hoping to obtain quantum mechanics, we make very mild assumptions:

\vspace{0.5cm}

\noindent 
0.) We shall make very mild assumptions, mostly mathematically
  almost always
  assumed by physicists. 
The assumptions are about the ``distributional'' $P$ such as continuity, differentiability, 
Taylor expandability, and that sort of things.

\vspace{0.5cm}

\noindent
1.) In addition we make a little less general assumption: $P$ is exponentially
  strongly varying,
  as if of the form
  \begin{eqnarray}
    P[q] &=& \exp\left(\frac{1}{\hbar} S_P[q] \right)     \label{P=exp(S_P/hbar)} \\
    \hbox{with }\frac{1}{\hbar} &\sim& \hbox{very large}. \nonumber
    \end{eqnarray}

\vspace{0.5cm}

\noindent
2.) Furthermore, we make a mild assumption: There exists weak ``interaction'' 
with roughly periodically moving  ``clock''.

\vspace{1cm}

Even almost empty assumptions and formulations may have drastic implications. 
Our formulation is so general that it also would accept a theory in which one has laws for 
what shall happen at some moment of time. 
It allows the future to be guiding for what happens or the past, as it seems to be in reality. 
If we wish, we could impose that every initial conditions would be equally likely; 
but in reality we have some ideas about the initial state (big bang, inflation, etc.). 
But making $P[q]$ or $\frac{1}{\hbar} S_P[q]= \ln(P[q])$  some nice smooth function 
might guide us towards getting such ``initial state predictions'' 
not coming from a single moment but being some compromise coming in a bit at all times.

\vspace{1cm}

\section{Phenomenological example of the specific form of $S_P$}

\subsection{Our formalism determines a favored path}

In functional integrals for quantum mechanics we have an action in the exponent 
\begin{eqnarray}
\hbox{``Functional integral''} 
&=& \int \exp\left(\frac{i}{\hbar} S[q] \right){\cal D}q.
\end{eqnarray}
The introduction of $S_P$ in Eq.(\ref{P=exp(S_P/hbar)}) tells us that 
it is not $S_P$ but $-iS_P$ that corresponds to an action $S[q]$. 
We can put in our own favorite action for $-iS_P[q]=S[q]$ and get our own equations of motion, 
but let us consider some system of particles as a typical example of the specific form of $S_P$: 
\begin{eqnarray}
    -i S_P[q]&=& \int L(\dot{q}(t), q(t)) dt \nonumber \\
    &=& \int (K(\dot{q}) - V(q(t))) dt, 
\end{eqnarray}
where $K(\dot{q})$ and $V(q(t))$ could be usual kinetic 
and potential energies respectively, say e.g., 
\begin{eqnarray}
  K(\dot{q}) &=& \sum_i \frac{1}{2}m_i  \dot{q}_i^2, \\
  V(q(t))& =&\hbox{``potential''}\hbox{(that could have peaks and valleys etc.)}.
\end{eqnarray}

When one seeks the path (= history) $q_\text{max}$ with the highest probability $P[q_\text{max}]$, 
one gets that the variation for it there, i.e., the functional derivative of the action, is zero, 
and derives an equation of motion (classically at least).  
We note that an overall sign or a constant multiplying the whole action does not change 
the equation of motion. 
But the relative weight of different paths (= histories) is violently influenced. 
Thus, our formalism determines a favored path to be realized.

\vspace{1cm}

\subsection{Does the favorite path fit the cosmology?}

For simplicity, we restrict ourselves to the uppermost $V$-potential favored case. 
Then the best path stands on the highest mountain. 
But, if there is a so broad distribution of path around the one with 
very highest $S_P[q]$ that they cannot all just stand on the peak, 
then there will be a flow down. 
Among the flow down paths the most favorable one for getting high $S_P$ would 
go up to another peak quickly.

\begin{figure}
\centering
\includegraphics[scale= 0.5]{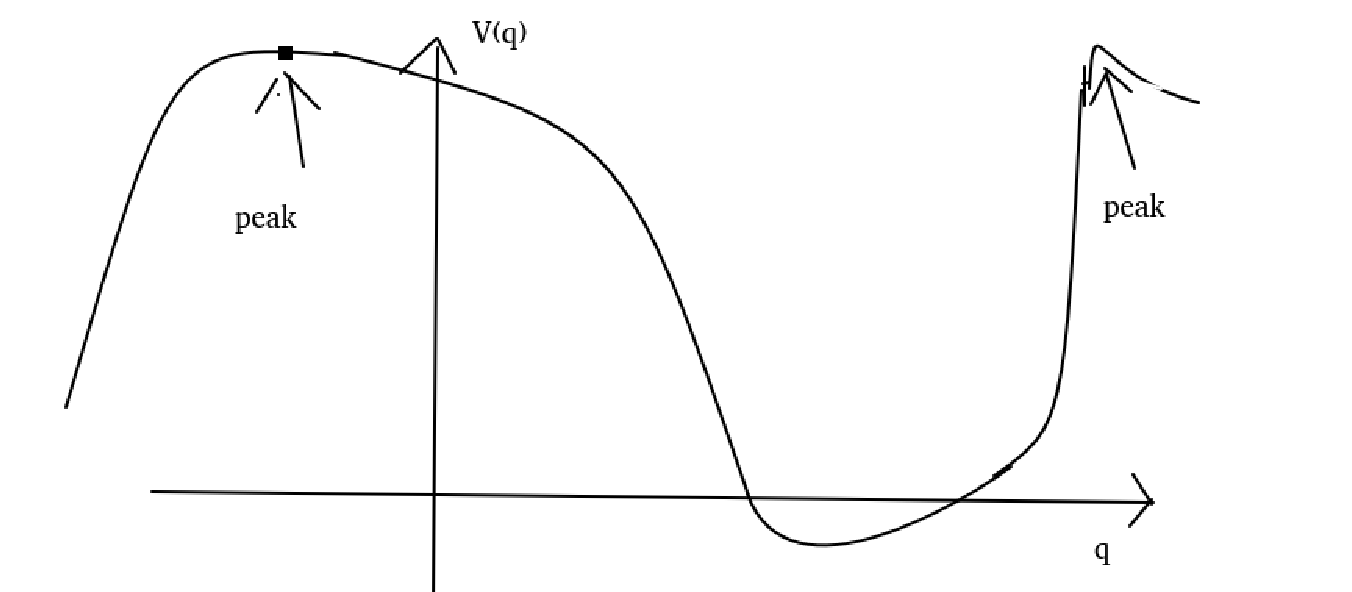}
\caption{Inflaton potential with two peaks}
\label{figure:inflaton_potential}
\end{figure}

\vspace{1cm}

Let us consider the behavior of the inflaton field in the inflation universe model by 
supposing a generic potential as drawn in Fig.\ref{figure:inflaton_potential} for the inflaton potential. 
We discuss it as follows:
\begin{itemize}
 \item{Waiting on an almost highest peak till it falls down by accident.}

The inflaton field is standing on a peak in the potential so long that 
    the physicists consider the famous long-staying problem: ``the slow roll problem''.

\item{It seeks quickly to find up to another similar peak to convert kinetic energy to potential energy.}

The Universe did after inflation expand with an enormous Hubble-Lemaitre
    constant, meaning that it brought quickly massive or massless particles away
    from each other, so that, in Newtonian gravity say, the gravitational potential energy
    should begin to raise as quickly as possible.

\item{It should stay again long on the next peak.}

The Hubble-Lemaitre has slowed down and the time scale of 
the development is now huge, compared to the one in the beginning (just after inflation stopped). 

\end{itemize}
We speculate that the above picture could be one of solutions to the slow roll problem. 
See also Ref.~\cite{Nielsen:2024iln}.

\vspace{1cm}

\section{Introducing a beating ``clock''}

In this section we discuss mainly how interference, which is one of the important properties of quantum mechanics, could be realized by considering a beating ``clock'' in our formalism.

\subsection{A beating ``clock'' and interference}

We begin with considering a couple of important properties of quantum mechanics. 
They are summarized as follows:  

\begin{itemize}
\item The system/ the particle can be several places at a time.

We already have that in our formalism from the point of view of the path integral.

\item When it can go two (or more) ways, the probability is not just
    additive, but depending on each phase, it could be bigger or smaller. This is an interference. 

\end{itemize}

\noindent
We obtain a kind of interference by speculating a ``clock'' interacting with the system. 
In the following we discuss how quantum interference is realized in our formalism.

\begin{figure}
\centering
\includegraphics[scale= 0.5]{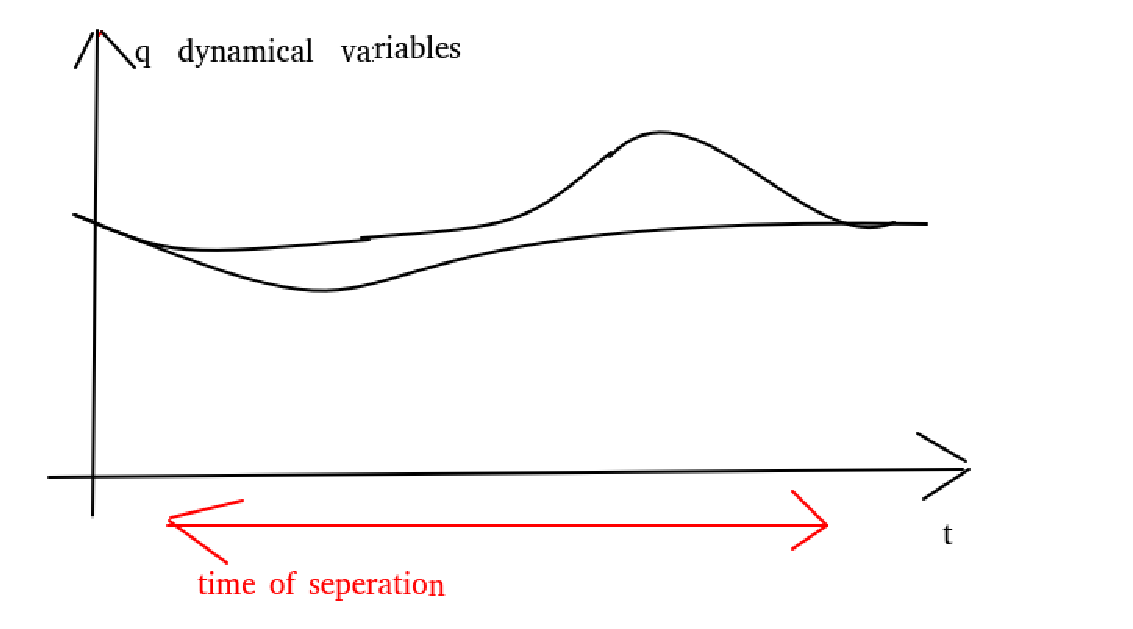} 
\caption{Separated paths}
\label{figure:separated_paths}
\end{figure}

\begin{figure}
\centering
\includegraphics[scale=0.5]{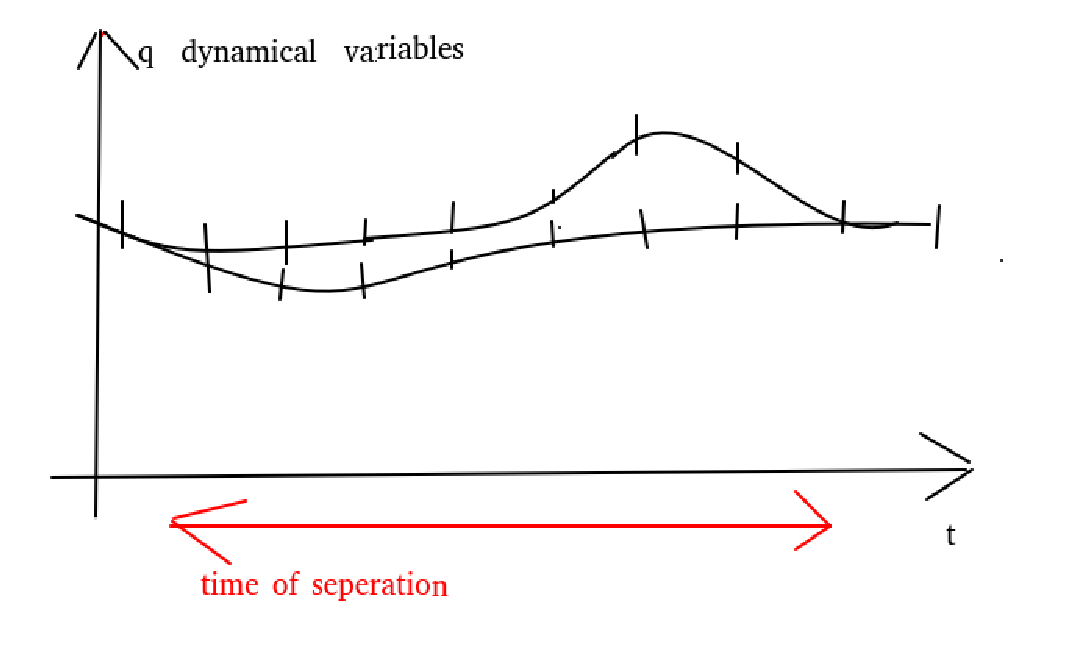}
\caption{A ``clock'' beating disconnected}
\label{figure:clock_beating_disconnected}
\end{figure}

\begin{figure}
\centering
\includegraphics[scale=0.5]{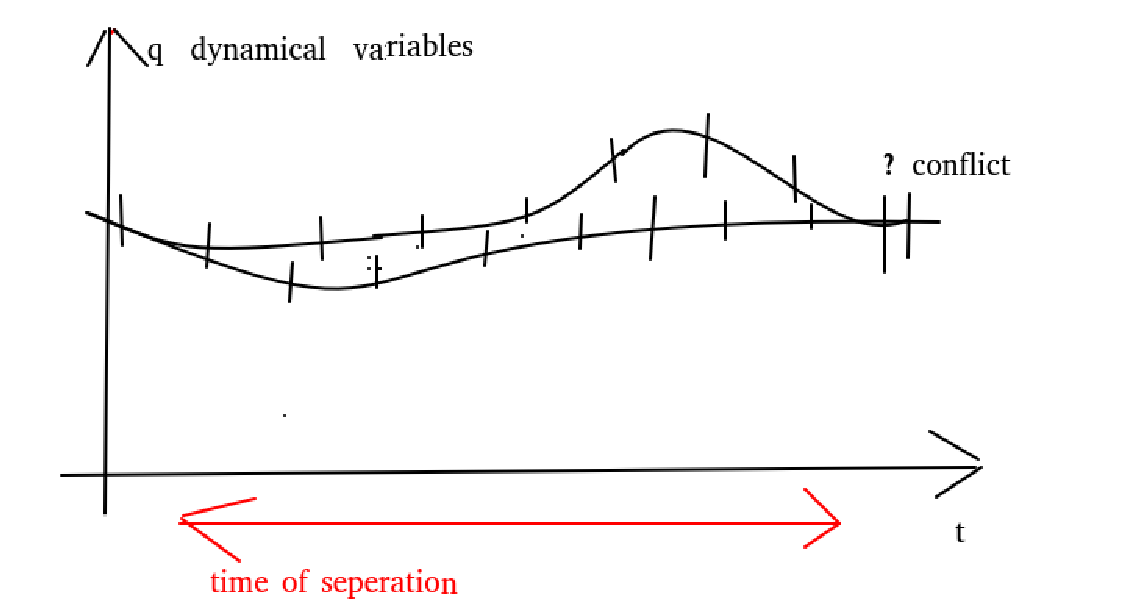}
\caption{Clocks beat differently on paths}
\label{figure:clock_beat_differently}
\end{figure}

Sometimes there are deviations from determinism, i.e., an optimal path 
could be separated into two paths as drawn in Fig.\ref{figure:separated_paths}. 
In addition, even if there is a beating ``clock'' on a path and it is disconnected as drawn in Fig.\ref{figure:clock_beating_disconnected}, it does not matter. 
But, if the ``clock'' goes faster on one of the separated tracks than 
on the other one, as drawn in Fig.\ref{figure:clock_beat_differently}, what happens? 
Really, there could be no separation then. 
Next, what happens in case of different beating rates in the two separate ways? 
Remembering that the deterministic (classical) tracks really represent  
narrow bundles of tracks - narrow because the coefficient $\frac{1}{\hbar}$ 
is very small -, one must imagine that small deviations from the best ``classical'' path 
just have a lower probability than the best classical track itself. 
So, at least small problems of the ``clock'' not beating consistently would lead to 
such probability decrease. 
But, if the difference of the times on the two tracks is just a shift 
by an integer number times the beating period, there would be no decrease.
This sounds like interference. 
We expect that it gives 
an imaginary term in the exponent of the functional integral, even though it might look tricky.
It would be interesting to calculate the suppression of probabilities from more or less 
consistent matching of the ``clock'' beating on various tracks.

\subsection{Local version of the ``clock'' and a charged particle}

To see that we hopefully are on the right track towards a realistic
model, we could make the ``clock'' be replaced by a separate little clock in
each point of space. The model with only {\it one} of these ``clock''s would not be truly local, 
so having clocks distributed all over space would be better
from such a  principle of locality. 
We have had in mind that 
these ``clock''s run so fast that we shall not be able to consider it for us
achievable knowledge where they are in their cyclic running. 
Now let us imagine a pattern of ``clock''s all over in space, and for simplicity, a system 
just with the degrees of freedom of a non-relativistic particle. 
If we say that we only have access to the difference in
progression along different paths in space time between the same two events
but not to how far the different clocks have reached at given moments, we
have strong similarity to the knowledge of electromagnetic fields, while not
knowing the gauge. 
In other words we propose to look at a system of infinitely many ``clock''s (one at every space point)  developing a little bit differently here and there, as representing a possibility of different 
behaviors just in correspondence with electromagnetic fields in space time.

It is not difficult to prove that, if the different phase deviations for the
many different loops of curves in Minkowski space shall be consistent in the
sense that, when one loop is composable from two, of course the phase
deviation for that loop must be the sum of those of the two components,
then we can find electromagnetic fields describing the phases for the
various loops.
The simplest realization of the just mentioned idea would be to simply
call the rate of running of the ``clock'' $clock(x)$ at position $x$ for
$A_0(x)$, meaning identifying it with the electric potential. Then we could
look at a gauge transformation in a purely electrostatic theory which
is  an addition of the same  constant to $A_0(x)$ at every point $x$ in space,
as a general increase in the running speeds of the small clocks.
Well, the idea we seek here to put forward is that there is hope for getting
the mysterious $i$ in quantum mechanics connected with clocks that really
are connected with the electric properties of the particle.
But if so, we might think that, if we had chargeless particles, which would typically be Majorana particles, 
then we should have real wave function for them. 
That is indeed true that single particle wave functions for Majorana fermions are real.

\vspace{1cm}

\section{The weak value and our general formalism}

\subsection{Weak value}

Our formalism with $P[q] = \exp(\frac{1}{\hbar} S_P[q])$ was originally inspired from 
and also is most easily connected to the formalism of quantum mechanics 
by means of the weak value:
\begin{eqnarray}
    O_\text{weak value}(t)&=& \frac{\langle f |O(t)|i\rangle}{\langle f |i\rangle}  \quad \hbox{(Heisenberg representation)}
\end{eqnarray}
where one is so to speak to know or put in some information on
  the initial state $|i\rangle$ given at the initial time $t_i$ and on the final state $|f\rangle$ given at 
the final time $t_f$. $O$ is an operator, say Hermitian. 
In the Schr{\"o}dinger representation, the weak value is expressed as 
\begin{eqnarray}
  O_\text{weak value}(t)&=& \frac{\langle f |\exp(- \frac{i}{\hbar} H(t_f-t))O\exp(-\frac{i}{\hbar} H(t-t_i))|i\rangle}{
    \langle f |\exp(- \frac{i}{\hbar}  H(t_f-t_i))|i\rangle}
\end{eqnarray}
where $H$ is a given Hamiltonian, and the states $|i\rangle$ and $|f\rangle$ are supposed to 
time-develop according to the Schr{\"o}dinger equation 
for a state $| \psi \rangle$: $i\hbar \frac{d}{dt} | \psi \rangle = H| \psi \rangle$.

Weak value is the most useful when we have complex action and in principle know even the future. 
We worked on such complex action theories, and the weak value formalism 
seemed very natural for the hypothesis we worked on that the action was {\it complex}. 
Indeed, in the Wenzel-Dirac-Feynman functional integral expression, the weak value of $q_i(t)$ 
is symbolically expressed as 
\begin{eqnarray}
    q_{i \; \text{weak value}}(t)&=& \frac{\int \exp(\frac{i}{\hbar}  S[q])*q_i(t) {\cal D}q}{
      \int \exp(\frac{i}{\hbar}  S[q]){\cal D}q} , \label{wv}
\end{eqnarray}
which is much simpler than the expression of the usual expectation value of $q_i(t)$.

Our great result was that we would get similar equations of motion as for real action, 
but only in addition to getting some predictions about the ``initial conditions''. 
We could say that complex action unites equations of motion with ``initial conditions''.

\subsection{Our maximizing overlap assumption and classical interpretation}

The choice of the final state $|f\rangle$ and initial state $|i\rangle$ will in most cases 
with complex action be determined by requiring that the absolute value of 
the transition amplitude from the initial state to the final state is maximized: 
\begin{eqnarray}
    |\langle f |i\rangle|& \text{is} & \text{maximal.} \label{maxoverlap} \quad \hbox{(Heisenberg)}\\
    |\langle f |\exp \left(- \frac{i}{\hbar}  H(t_f-t_i) \right)|i\rangle|& \text{is} & \text{maximal.} \quad \hbox{(Schr{\"o}dinger)}
\end{eqnarray}
In real action case we get from this (\ref{maxoverlap}) still an
  undetermined set of states but we get
\begin{eqnarray}
    |\langle f |i\rangle|_{max} &=& 1 \hbox{ for usual real action case}.\nonumber\\
    &&\hbox{(Heisenberg)}\nonumber\\
    |\langle f |\exp\left(- \frac{i}{\hbar}  H(t_f-t_i) \right)|i\rangle|_{max}&=& 1 \hbox{ for usual real action case}.
    \nonumber\\
    &&\hbox{(Schr{\"o}dinger)}\nonumber 
\end{eqnarray}

One would wonder how we can think classically in complex action and weak value. 
It is summarized as follows: 
\begin{itemize}
  \item With complex action, typically all that happens in universe at
    all times gets predestined (because it is a theory also for the
    ``initial conditions''). 

  \item In the large $\frac{1}{\hbar}$ approximation only one or very few
    classical paths are realized, and the one with highest probability
    wins.

  \item Paths have a few times where they split up into two or more.

  \item If we arrange by the ``clock'' story to make ``interference'', 
    it can modify the total probability of the path with the splitting that has the ``interference'' correction.
  \end{itemize}

\begin{itemize}
  \item Such ``interference'' corrections may cause an otherwise
    winning path to get beaten by a slightly less probable competing path.
    (Presumably we shall imagine a sample of near competitors clear to
    take over if a path gets too much destructive interference.) 

  \item For a dominant path, i.e., in the classical approximation, the weak value 
for an operator is simply the value of the corresponding dynamical variable at the time $t$.

  \item So we can look at the weak value as just a way to extract the
    classical path, which is determined by our imaginary part of the action. 

\item Different (bunches of) paths have different probabilities; a path with highest probability is 
the likeliest to be realized as our history.

\end{itemize}

\begin{figure}
\centering
\includegraphics[scale=0.5]{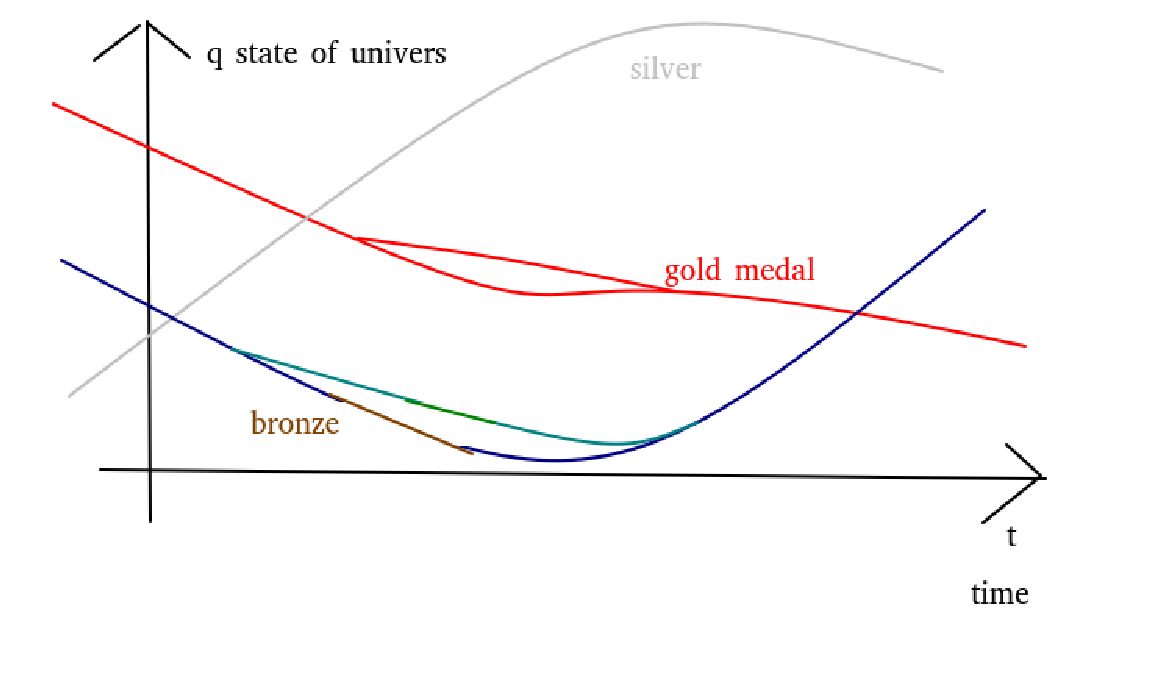}
\caption{Competing paths}
\label{figure:competing paths}
\end{figure}
%

\subsection{Interpreting the effect of  the ``clock'' as giving us the weak value with complex path integral}

  Now we want to formulate the result of our general formulation including
  the ``clock'' to lead to that expectation value of one of the $q$ variables
  or a combination of them being the weak value
\begin{eqnarray}
    q_{i \; \text{weak value}}(t) = \langle q_i(t)\rangle&=& \frac{\int \exp(\frac{i}{\hbar} S[q])*q_i(t) {\cal D}q}{
      \int \exp(\frac{i}{\hbar} S[q]){\cal D}q}. \label{wv2}
\end{eqnarray}
If the probability density $P[q]$ contains much information on the
  initial and final states, it will not be so serious to ignore the
  boundary conditions, because this information will then be transfered into
  $S[q]$ used in this formula (\ref{wv}).

We compare our general formulation with the weak value in quantum mechanics. 
An average of one of the $q$-variables $q_i(t)$ at time $t$ is expressed as 
\begin{eqnarray}
q_i(t)_\text{weak value}=\langle q_i(t)\rangle 
&=& \frac{\int \exp(\frac{i}{\hbar} S[q])*q_i(t) {\cal D}q}{\int \exp(\frac{i}{\hbar} S[q]){\cal D}q}, \\
\langle q_i(t)\rangle_\text{our formalism} &=& \frac{\int P[q] q_i(t){\cal D}q}{\int P[q]{\cal D}q}, 
\end{eqnarray}
where the denominator $\int P[q]{\cal D}q $ is just a normalization. This will not be needed 
if $P[q] $ is already normalized.

We see an important difference: the weak value consists of {\it complex} integrals, 
while in our formalism everything is {\it real} numbers. 
The similarity gets even bigger formally when we remember that we want to 
assume as a helping assumption that $S_P[q]$ is supposed rather smooth, 
so that the form
  \begin{eqnarray}
    P[q]&=& \exp\left(\frac{1}{\hbar} S_P[q] \right)
    \end{eqnarray}
is called for. Remember also that $q$ stands for a set of functions, 
so it really means what we would call a track, a path, or a history of the universe.
But, to make the agreement between the two expressions, we would need to 
provide our expression with an artificially invented phase.

\vspace{1cm}

  Remembering the $q$ variables are supposed real by themselves, we see that, 
if only one history or track $q$ dominates, then weak value becomes real 
for the different $q_i$ which are assumed real/ Hermitian as operators, 
because, for a single dominating path, there is a reality theorem for the weak value. 
Even if we have a history with some separation from time to time, 
but we ask for the weak value for $q_i(t)$ at a time outside the separation period, 
then the weak value for this $q_i(t)$ will be real.
Thus the weak value gives perfect description in classical case, meaning one track or history dominates.
\begin{figure}
\centering
\includegraphics[scale=0.5]{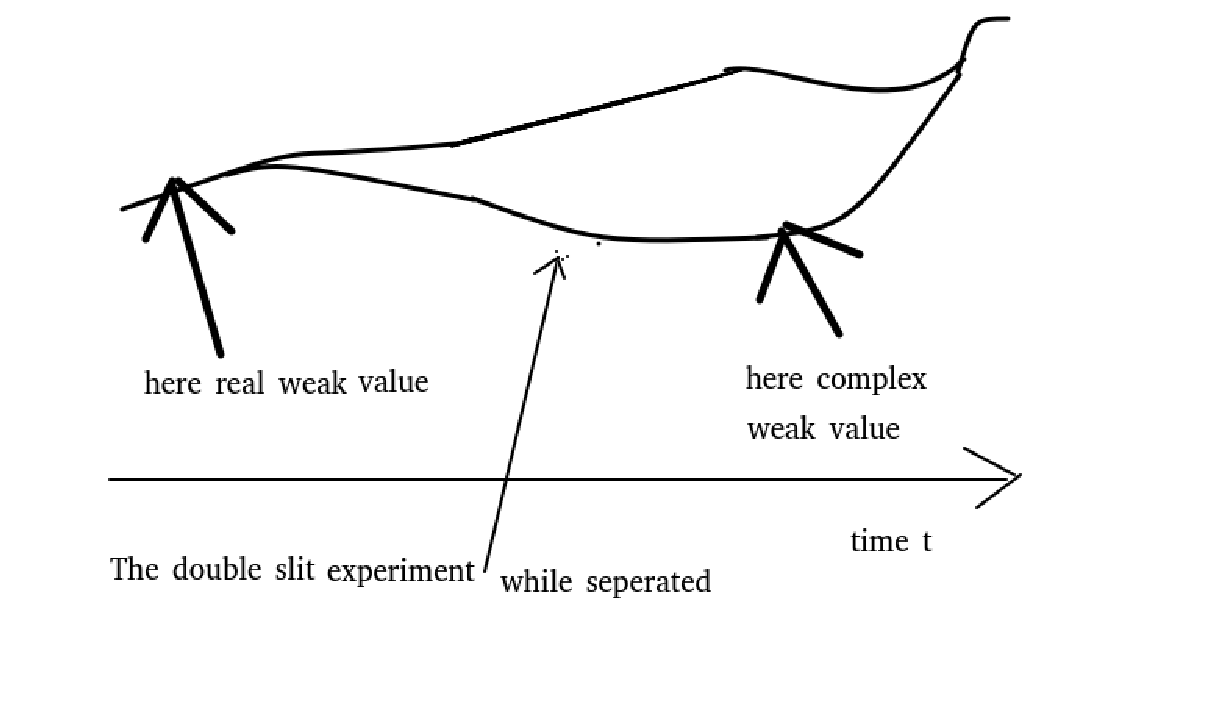}
\caption{Double slit experiment paths}
\label{figure:Double_slit_experiment_paths}
\end{figure}
If physicists make double slit experiments where a particle goes through 
two slits simultaneously, the weak value still gives averages, but 
now the average will usually be {\it complex}, as we know that 
asking a stupid question like ``Through which slit did the particle go?'' in an interference experiment 
gives a stupid / complex answer. 
On the other hand, our formalism - before we modify it very artificially - cannot give complex answer, 
because it is made so, as if it never heard about complex numbers. 
Therefore, our formalism gives a priori real numbers even for the average while the particle is passing through the double slit experiment.

\vspace{1cm}

\subsection{Adding a phase to $P(q)$ in our formalism}

To have our general formalism match the weak value, we have to provide our expression 
with an artificially invented phase, i.e., we need to add a phase formally to our $P[q]$ 
to make it look like the integrand in the Feynman-Dirac-Wentzel functional integral. 
For this purpose, we define a ``clock'' delay ratio for any path/history $q$ at any moment of time $t$: 
\begin{eqnarray}
    \delta [q,t] &=& \hbox{``time delay ratio in period''} \nonumber \\
    &=& \frac{c_\text{~standard}[q,t]- c[q,t]}{\hbox{period}},  
\end{eqnarray}
where\footnote{$\delta$ and the $c$'s are functionals of $q$, but functions of $t$.}
\begin{eqnarray}
&& c_\text{~standard}[q,t] = t, \\
&& c[q,t] = \text{the stand of the ``clock''on path $q$ at time $t$}. 
\end{eqnarray}
Then we are first suggested to put 
\begin{equation}
iS[q] = S_P[q] + i\delta [q, t].  \label{reltowv}
\end{equation}
Remember that $S_p[q]$ is the logarithm of the probability in our model formulation and thus 
of course real, while we like the complex action theory (CAT), 
in which $S[q]$ is complex - while in usual theory real -. 
By putting in the real $\delta[q,t]$ with an $i$, we obtain the right hand side 
of (\ref{reltowv}) being complex with both a real and an imaginary part.

We imagine the ``clock'' to have very short ``period'', so the time of the clock is not so important 
if we use some average period or one about the time $t$. 
In fact, we expect the probability for a path with a split time, as represented by 
a double slit experiment, to be somewhat reduced, 
because of the interaction of the other degrees of freedom with the ``clock'' 
by means of the action $S_P[q]$ that is purely imaginary from a usual point of view in our formulation, 
i.e., because of the not matching of the ``clock'' and the rest of the system.
We hope to have our model give quantum mechanics such that 
this reduction turns out to be equal to the effect of having the phase addition as we suggested. 
So, in order not to have it doubled, we improve our suggestion to
\begin{eqnarray}
    iS[q] &=& S_P[q]|_\text{with ``clock'' removed}+i\delta[q,t]. 
\end{eqnarray}
Here $S_P[q]_\text{with ``clock'' removed}$ is a modified $S_P[q]$ in our model where 
we hope that the ``clock'' is removed, and so only the rest is left.

We hope to calculate that removing ``clock'' and adding the phase just cancel each other. 
For two numerically equally high probability paths during the separation seems likely 
by the probability proportional to 
$\vert 1 + e^{i\Delta} \vert^2 \propto \cos^2(\frac{1}{2} \Delta)$, where 
$\Delta$ is an average of $\delta[q,t]$ over separation. 
We will argue for that, at least being right in the small deviation between the delay 
in the two separate paths case.

\vspace{1cm}

\section{Discussion}

We have put up a very general formalism, in which we may reproduce 
rather usual classical actions although it comes with the $i$ missing relative 
to the usual functional integral. 
But for the action in classical 
physics an overall sign as e.g. an $i$ does not matter. 
In the very general model just having probability density
$P[q]$  as a functional that we can adjust phenomenologically, describing the
probability density for all histories a priori still to be evaluated by the Taylor expansion and the like, 
we have one property of quantum mechanics already:
\begin{itemize}
\item The system/the world can go through different paths, so the state at a moment $t$ is not quite unique.

\item Only after assuming the exponent $S_P[q]$, when $P[q] = \exp(\frac{1}{\hbar} S_P[q])$ is very large, 
we obtain classical physics: only one path is realized with high probability. 
\end{itemize}

The weak value for a quantity (= dynamical variable) and the expectation value 
in our general formalism with $P[q] = \exp( \frac{1}{\hbar} S_P[q])$ only deviate by an $i$, 
as seen by comparing the weak value expression and our expectation value one:
\begin{eqnarray}
&& \langle O(q(t))\rangle 
= \frac{\int \exp(\frac{1}{\hbar} S_P[q])O(q(t)){\cal D}q}  {\int \exp(\frac{1}{\hbar} S_P[q]){\cal D}q} 
\quad \text{(Our average)}, \\
\nonumber\\
&& \langle O(q(t))\rangle_\text{weak value} 
= \frac{\int \exp(\frac{i}{\hbar} S[q]) O(q(t)){\cal D}q}{\int \exp(\frac{i}{\hbar} S[q]){\cal D}q} 
\quad \text{(The weak value)}, \nonumber \\
\end{eqnarray}
provided we identify the ``actions'' as follows:
\begin{eqnarray}
    S_P &=& i S.  
\end{eqnarray}
We note that classical equations of motion are not sensitive to this $i$. 
The classical approximation or equations of motions for both of them become the same 
in spite of the $i$ separating them. 
However, our general formalism has a very strong  - and
    not shared by the weak value with the $i$ - prediction about the
    {\it initial  state conditions.}

The features being favored by our formalism may be matched 
with the very strongest features of cosmology: slow inflation, huge expansion 
in the beginning after the ``reheating'', and much slower expansion in the long run. 
But we have the problem that our model tends to make a decision about
the initial conditions. 
We proposed  a way to - by not quite finished calculations - obtain a relation 
between our general formulation and the weak value formulation of quantum mechanics, 
especially in the case of an action being complex, i.e., our complex action theory. 
In the short run, we could easily arrange by choosing $S_P$ that 
one would not notice in short terms the tendency of the model to give information 
on the initial conditions (and possibly also on the future being selected). 
But the interference needed an extra story: the ``clock''. 
We shall hopefully prove that with this``clock'' we can obtain the usual quantum mechanics 
with its mysterious complex numbers. 
The density operator that we introduced in the future-included CAT~\cite{Nagao:2022jam} 
would be useful for the study. 
As a by-product - but may be most interesting - we found that the path favored in probability 
had some similarities in general properties of escaping as fast as possible 
not being at the maximal potential energy. 
This could be interpreted as a model behind slow roll and fast Lemaitre-Hubble expansion in the beginning.

\section*{Acknowledgments}

H.B.N. is grateful to NBI for allowing him to work there as emeritus. 
This work was supported by JSPS KAKENHI Grant Number JP21K03381, and partly 
accomplished during K.N.'s sabbatical stay in Copenhagen. 
He would like to thank the members and visitors of NBI for their kind hospitality 
and Klara Pavicic for her various kind arrangements and consideration during his visits to Copenhagen.  
Furthermore, the authors would like to thank Astri Kleppe for her useful discussion at the early 
stage of this work, and the organizers of Bled workshop 2024 for their kind hospitality. 




\begin{thebibliography}{9}
\bibitem{RDAstri} There is a home page by Astri Kleppe having collected
    some works on Random Dynamics as well as some original comments.
\end{thebibliography}
\end{document}